\documentstyle[11pt,dunk2001_asp,twoside]{article}

\def\oversim#1#2{\lower0.5pt\vbox{\baselineskip0pt \lineskip-0.5pt
     \ialign{$\mathsurround0pt #1\hfil##\hfil$\crcr#2\crcr\sim\crcr}}}

\def\lesssim{\mathrel{\mathpalette\oversim<}}   

\def\edcomment#1{\iffalse\marginpar{\raggedright\sl#1\/}\else\relax\fi}
\reversemarginpar

\begin{document}

\title{Formation of Galactic Disks}

\author{S. Michael Fall}

\affil{Space Telescope Science Institute, 
       3700 San Martin Drive,
       Baltimore, MD 21218, USA}

\begin{abstract}
We review progress in understanding the formation of galactic disks 
in the standard cosmogonic scenario involving gravitational clustering
of baryons and dark matter and dissipative collapse of the baryons. 
This scenario accounts remarkably well for the observed properties 
of galactic disks if they have retained most of the specific angular 
momentum they acquired by tidal torques.
Early simulations, which included cooling of the gas but not star
formation and the associated feedback, indicated instead that most 
of the angular momentum of the baryons would be transferred to the 
dark matter.
Recent simulations indicate that this angular-momentum problem can 
be solved partially, and in some cases entirely, by feedback and 
other effects.
\end{abstract}

\section{Introduction}

Two key ingredients in the formation of galactic disks are
dissipation and rotation.
Dissipation by radiative cooling causes the gas in a protogalaxy 
to collapse inward; rotation then halts the collapse in the 
directions perpendicular but not parallel to the overall 
angular-momentum vector, resulting in a thin centrifugally 
supported disk.
Other processes likely to play some role in the formation and 
subsequent evolution of galactic disks include dynamical friction, 
internal torques, kinematic viscosity, mergers, and star formation 
and the associated heating and stirring of the gas (feedback).
In this article, we consider the general cosmogonic scenario in 
which the extended halos of galaxies form hierarchically by the 
gravitational clustering of non-dissipative dark matter, and the 
luminous components of galaxies form by a combination of the 
gravitational clustering and dissipative collapse of baryons, 
as proposed by White \& Rees (1978).
The formation of galactic disks and the origin of their rotation
in this scenario were first studied by Fall \& Efstathiou (1980).
Most of the currently popular models of galaxy formation, including 
all variants of the cold and warm dark matter (CDM and WDM) models, 
are specific versions of this general scenario.

\section{Tidal Torques}

The formation of objects by gravitational clustering automatically
endows them with some angular momentum (Peebles 1969).
This is because the (proto)objects initially have irregular shapes 
and hence non-zero quadrupole moments. 
They therefore exert tidal torques on each other, with random strengths 
and directions. 
The objects acquire most of their angular momentum in the translinear
regime, when their density contrasts are appreciable ($\delta\rho/\rho
\sim 1$), but before they reach their maximum sizes and begin to 
contract. 
The rotation induced by tidal torques is usually quantified in terms 
of the dimensionless spin parameter
$$
\lambda \equiv J |E|^{1/2} G^{-1} M^{-5/2},   
\eqno(1)
$$
where $J$, $E$, and $M$ are the total angular momentum, energy, and
mass of the object, and $G$ is the gravitational constant.
Rapidly rotating objects, such as disks, have $\lambda \sim 1$,
whereas slowly rotating objects, such as spheroids, have $\lambda
\ll 1$. 

Cosmological $N$-body simulations have revealed that the distribution
of spins induced by gravitational clustering alone (i.e., without
dissipation) is approximately lognormal:
$$
p(\lambda)d\lambda \propto \exp \bigg\{ - \frac{1}{2} 
  \bigg[ \frac{\ln (\lambda/\lambda_m)}{\sigma(\ln \lambda)} \bigg]^2 
  \bigg\} d \ln\lambda,
\eqno(2)
$$
with
$$
\lambda_m \approx 0.05 \qquad {\rm and} \qquad
\sigma(\ln \lambda) \approx 0.5.
\eqno(3)
$$
Most objects rotate slowly although there is a wide range of spins.
This distribution appears to be nearly universal in the sense that 
it has little or no dependence on the cosmological parameters, the 
initial spectrum of density perturbation, or the masses and densities
of the objects (Barnes \& Efstathiou 1987; Zurek, Quinn, \& Salmon 
1988; Warren et al.\ 1992; Cole \& Lacey 1996).
These results are very useful, even if they are not yet fully 
understood theoretically.

\section{Analytical Collapse Model}

We can relate the factor by which the baryons collapse in the 
radial direction before they reach centrifugal balance in a disk 
to the initial rotation of a protogalaxy as follows (Fall 1983).
For simplicity, we consider a galaxy with a luminous disk (D) 
and a dark halo (H) but not a luminous spheroid. 
Furthermore, we assume that the baryons destined to become the 
disk receive the same tidal torques as the dark matter before
much dissipation occurs and that during and after the collapse
the total specific angular momentum of each component is
conserved: 
$$
J_D/M_D = J_H/M_H.
\eqno(4)
$$
In other words, we assume for now that angular momentum is not
transferred between the disk and the halo, or if it is, that
this is accompanied by the transfer of enough mass (outflow or 
inflow) that equation~(4) is still satisfied. 

We approximate the disk by an exponential model with a scale 
radius $\alpha^{-1}$ (sometimes denoted by $R_d$) and the halo 
by a singular isothermal sphere with a circular velocity $v_c$ 
and a truncation radius $r_t$. Then, neglecting the self-gravity 
of the disk, we have 
$$
J_D/M_D = 2 v_c \alpha^{-1},
\eqno(5)
$$
$$
J_H/M_H = \sqrt 2 \lambda v_c r_t.
\eqno(6)
$$
Equating these gives a very simple relation between the collapse 
factor $\alpha r_t$ of the baryons in the disk and the spin 
parameter $\lambda$ of the halo:
$$
\alpha r_t = \sqrt 2 /\lambda.
\eqno(7)
$$
This implies $\alpha r_t \sim 30$ for $\lambda \sim 0.05$ and 
hence $r_t \sim 100$~kpc for a typical spiral galaxy like the 
Milky Way (with $\alpha^{-1} \approx 3$~kpc).
The collapse factor is larger in halos with smaller spin
parameters and smaller in halos with larger spin parameters.
The collapse factor would also be larger if the baryons were 
to lose some of their specific angular momentum.

Equation~(7) is an excellent approximation even when the halo 
has a finite core radius and when the self-gravity of the 
disk is included (Fall \& Efstathiou 1980; see their Fig.~3).
It is also a good approximation if the halo has an NFW profile 
(Navarro, Frenk, \& White 1996, 1997) that is later deformed by 
the (adiabatic) contraction of the disk within it (Mo, Mao, \& 
White 1998).
The radius $r_t$ of the halo in equation~(7) should be the 
one within which the baryons have collapsed onto the disk.
This may be compared and contrasted with the radius $r_{200}$ 
within which the mean density of the protogalaxy is 200 times 
the critical (closure) density.
The latter is near the transition between the virialized and 
infalling parts of the halo (Cole \& Lacey 1996). 
In recent work in this field, it has been customary to identify 
$r_t$ with $r_{200}$.
This assumption, however, does not have much physical justification
beyond the constraint $r_t \lesssim r_{200}$.
It is likely that $r_t$ depends on cooling, heating, and other 
non-gravitational processes at least as much as it depends on 
the gravitational clustering that determines $r_{200}$.
Thus, we have no guarantee that $r_t$ and $r_{200}$ will be
equal or even proportional to each other.
(Anyone who doubts this should compare the cooling and virial
radii in clusters of galaxies.)

\section{Scaling Relations}

The properties of the halos that form in a hierarchy by 
gravitational clustering obey some simple scaling relations.
We can combine these with the relation between the collapse 
factor and the spin parameter to derive the corresponding
scaling relations for the properties of galactic disks as 
follows (Fall 1983).
We approximate the relation between the typical masses $M_H$
of the halos (within $r_t$) and their circular velocities $v_c$ 
by a power law, $M_H \propto v^k_c$. 
The index $k$ depends in general on the initial spectrum of 
density perturbations, the cosmological parameters, the range 
of masses considered, and the relation between $r_t$ and 
$r_{200}$.
Simple arguments based on the different formation times
of objects with different mean densities give $k = 12/(1-n)$ 
for $(\delta\rho/\rho)_{\rm rms} \propto M^{-(3+n)/6}$ (White 
\& Rees 1978).
This implies $k \approx 4$ for the effective index $n \approx -2$ 
of the CDM spectrum on galactic scales (Blumenthal et al.\ 1984).
$N$-body simulations indicate instead $k \approx 3$ for the 
NFW halos in CDM cosmogonies (Navarro et al.\ 1997).
This index is appropriate for $r_t = r_{200}$ and hence for 
perfectly synchronous formation.
In view of the unknown initial spectrum (index $n$) and 
unknown relation between $r_t$ and $r_{200}$, we regard 
$k$ as a parameter with some theoretical uncertainty.

The scaling relation for the halos can be reexpressed in terms of 
the typical luminosities and circular velocities of the disks as
$$
L_D \propto (f_D/\Upsilon_D) v_c^k,
\eqno(8)
$$
where $f_D \equiv M_D/M_H$ is the ratio of masses of the disks
and halos, and $\Upsilon_D \equiv M_D/L_D$ is the mass-to-light 
ratio in the disks.
This may be compared with the observed Tully-Fisher (1977) relation, 
$L_D \propto v_c^l$, the index of which varies from $l \approx 3$ 
in the $B$ band (0.44 $\mu$m) to $l \approx 4$ in the $K$ band 
(2.2 $\mu$m) (Verheijen 2001).
Most of this variation can be explained by a dependence of
$\Upsilon_D$ on $v_c$, indicated by the observed correlation 
between the colors and the luminosities of galaxies.
When this effect, including the mass of interstellar gas, is taken 
into account, the index of the baryonic Tully-Fisher relation is 
found to be $k = 3.5 \pm 0.4$ for $f_D = {\rm const}$ (Bell \& 
de Jong 2001).

The central surface brightness of an exponential disk is given
by $I_0 = \alpha^2 L_D/2\pi$. 
Using equations~(7) and (8) and the relation $v_c^2 = G M_H/r_t$, 
we can rewrite this in the form
$$
I_0 \propto \lambda^{-2} v_c^{l-2k+4} \propto v_c^{l-2k+4},
\eqno(9)
$$
where all of the dependence on $f_D$ and $\Upsilon_D$ is contained
in the factor $v_c^{l-k}$.
The second proportionality holds because the spin parameters of 
the halos are statistically independent of their other properties.
For $k \approx 3.5$, we expect $I_0 \approx {\rm const}$ in the 
$B$ band ($l \approx 3$) and $I_0 \propto v_c$ in the $K$ band
($l \approx 4$). 
The first of these is the same as the original Freeman (1970) 
relation in the $B$ band. 
It would be interesting to search for the predicted correlation
in the $K$ band.

The scaling relations above indicate how the typical luminosities
and surface brightnesses of galactic disks depend on their circular 
velocities.
The dispersions about these relations are determined in part by 
the dispersion in the spin parameter.
We expect the $L_D$-$v_c$ correlation to have relatively little 
scatter because it is independent of $\lambda$ [equation~(8)], 
while we expect the $I_0$-$v_c$ correlation to have a great deal
of scatter because it includes the factor $\lambda^{-2}$
[equation~(9)].
These trends are qualitatively consistent with the observed
Tully-Fisher and Freeman relations. 
The distribution of $I_0$ at fixed $v_c$ can be derived from 
equation~(2) for $p(\lambda)$ and equation~(7) for $\alpha r_t$ 
(Dalcanton, Spergel, \& Summers 1997; Mo et al.\ 1998; Weil, Eke, 
\& Efstathiou 1998). 
These results can then be combined with the luminosity function
of galaxies to derive the joint distributions of $I_0$ and
$\alpha^{-1}$ (Dalcanton et al.\ 1997) and $L_D$ and $\alpha^{-1}$
(de Jong \& Lacey 2000).
In all these studies, the observed distributions are reproduced 
quite well provided the coefficient in equation~(7) is reasonably
close to $\sqrt 2$, i.e., provided the disks have retained most 
of the specific angular momentum they acquired by tidal torques.

\section{Origins of Exponential Disks}

The results above are based on the assumption that the {\it total} 
specific angular momenta in the disks are the same as those in their 
halos [equation~(4)]. 
We may refer to this as the {\it weak} form of the assumption of 
angular-momentum conservation.
In addition, it is sometimes supposed that the {\it distributions}
of specific angular momentum in the disks are the same as those in
their halos. 
This is usually expressed in terms of the fractions of mass with
specific angular momentum below $h = R v_{\phi}$ in the two
components:
$$
M_D(h)/M_D = M_H(h)/M_H.
\eqno(10)
$$
We may refer to this as the {\it strong} form of the assumption 
of angular-momentum conservation.
It is analogous to Mestel's (1963) hypothesis that $M_D(h)$
would be conserved in the collapse of a one-component protogalaxy 
(made before dark halos were discovered).
Note that equation~(10) implies equation~(4), but the converse 
is not true in general.
It is possible to change $M(h)/M$ in either component without 
affecting the corresponding $J/M$.

Several authors have pointed out that the distribution $M_D(h)/M_D$
for an exponential disk with a flat rotation curve is similar
to that for a uniform sphere with a constant angular velocity 
(Gunn 1982; van der Kruit 1987; Dalcanton et al.\ 1997).
This in turn resembles the distribution $M_H(h)/M_H$ caused by 
tidal torques (Barnes \& Efstathiou 1987; Quinn \& Zurek 1988), 
although there has been a recent tendency to emphasize the 
differences rather than the similarities (Firmani \& Avila-Reese 
2000; Bullock et al.\ 2001; van den Bosch 2001).
Thus, the strong form of the assumption of angular-momentum
conservation provides a possible explanation for the fact that 
the radial profiles of most galactic disks are approximately 
exponential.
The main concern here is that several processes could alter 
$M_D(h)/M_D$ and hence the radial profiles of the disks over 
their lifetimes.
These include the torques exerted by non-axisymmetric features 
in the disks (bars and spiral arms), the non-conservation of
angular momentum as gas flows though shocks in spiral arms, and 
the exchange of angular momentum in collisions between clouds, 
not to mention galactic outflows, fountains, and mergers.
It seems likely that these processes play some role in
determining the present distributions of specific angular 
momentum in galactic disks.

Lin \& Pringle (1987) showed that the radial distributions of the 
stars in galactic disks would evolve toward exponential-like 
profiles if two conditions were met. 
(1) The net effect of the processes that redistribute angular 
momentum in the interstellar gas can be described by an effective 
viscosity $\nu_{\rm eff}$.
(2) The associated timescale, $t_{\nu} = R^2/\nu_{\rm eff}$,
is about the same as the timescale for star formation, $t_* = 
\Sigma_g/\dot\Sigma_s$.
This is plausible because both the redistribution of angular
momentum and the formation of stars may be regulated by
instabilities in the disks, including bars and spiral arms,
although we lack a full understanding of just how or even
whether this would actually happen. 
The Lin-Pringle model has been explored further, including its
chemical evolution, by several authors (Clarke 1989; Yoshii \&
Sommer-Larsen 1989; Sommer-Larsen \& Yoshii 1989, 1990).
This model raises the possibility that the ubiquitous exponential 
profile is the result of viscous processes operating in the disks 
after they formed rather than tidal torques acting on the galaxies
while they formed.
More likely, both are involved: the exponential profile is first 
established by external torques and is then reinforced by internal 
viscosity.

\section{The Angular-Momentum Problem}

About a decade ago, it became possible to simulate the hierarchical 
formation of galaxies with both dark matter and baryons by a 
combination of $N$-body and hydrodynamical models.
The early simulations included the radiative cooling of the 
gas but not the formation of stars and the associated feedback 
(heating and stirring of the gas).
The results were very different from the analytical models
described above (Navarro \& Benz 1991; Navarro \& White 1994).
The baryonic objects that formed in these simulations were 
ellipsoidal in shape and an order of magnitude smaller than 
galactic disks of the same scaled mass. 
Thus, they resembled the spheroids more than the disks of real 
galaxies.
The reason for this is that the gas cooled quickly and collapsed 
into dense subunits within the halos.
A combination of dynamical friction and gravitational torques 
within the halos then transferred most of the orbital angular 
momentum of the baryons to the dark matter, causing the subunits 
to sink toward the centers of the protogalaxies, where they then 
merged.
(For a prescient discussion of how these processes might determine 
the different properties of galactic spheroids and disks, see Zurek
et al.\ 1988.)
The discrepancy between the baryonic objects produced in such 
simulations and real galactic disks has become known as the 
angular-momentum problem of galaxy formation.
It is so severe that it is sometimes referred to as a crisis or 
catastrophe.

Steinmetz \& Navarro (1999) have explored this and related issues 
with a series of high-resolution simulations. 
They have tried to alleviate the angular-momentum problem by 
including some feedback in their simulations but find this makes 
little difference to the outcome.
This conclusion, however, depends on their particular prescription 
for feedback.
Steinmetz \& Navarro assume that the gas is heated only in the 
immediate vicinity of ongoing star formation.
Then, since the gas is very dense at these locations, the energy 
input is quickly radiated away, before it can influence the motions 
or the thermodynamic state of the gas at other locations.
Steinmetz \& Navarro (1999) have also found that the zero-point 
of the relation between luminosity and circular velocity 
in their simulations differs significantly from that of the 
observed Tully-Fisher relation, especially in the Einstein-de 
Sitter cosmological model ($\Omega_M = 1$, $\Omega_{\Lambda} 
= 0$) with CDM.
More recently, Eke, Navarro, \& Steinmetz (2001) have shown that 
this problem is reduced substantially or solved completely in the
concordance cosmological model ($\Omega_M = 0.3$, $\Omega_{\Lambda} 
= 0.7$) with CDM or WDM.

\section{Possible Solutions}

It is now widely believed that the solution of the angular-momentum
problem must be found in some extreme form of feedback that prevents
the baryons from collapsing until after the violent relaxation in 
their halos is complete.
The baryons would then collapse within relatively smooth halos, 
with little quadrupole coupling, and hence would retain most 
of their specific angular momentum, as was originally, although
perhaps naively, envisaged (Fall \& Efstathiou 1980).
That this idea goes a long way toward solving the angular-momentum
problem has been demonstrated explicitly in a number of recent
simulations (Weil et al.\ 1998; Sommer-Larsen, Gelato, \& Vedel
1999; Eke, Efstathiou, \& Wright 2000; Thacker \& Couchman 2001).
In the Weil et al.\ and Eke et al.\ simulations, the gas evolves
adiabatically until the redshift $z = 1$ and is then allowed to 
cool radiatively and collapse.
In the Sommer-Larsen et al.\ and Thacker \& Couchman simulations,
the feedback is driven by local star formation but is extensive 
in both time and position (in contrast to the Steinmetz \& Navarro
prescription).
The common result of these studies is that the disks retain much 
more of their specific angular momentum. 

Several other effects also help.
The transfer of angular momentum is less severe in simulations
with warm dark matter than in simulations with cold dark matter 
(Sommer-Larsen \& Dolgov 2001).
The reason for this is that the halos in WDM simulations have
much less substructure and hence less dynamical friction and
internal torques than the halos in CDM simulations. 
The transfer of angular momentum is also reduced in simulations 
with a cosmological constant (Eke et al.\ 2000).
This happens because protogalaxies acquire their angular momenta
earlier, before the gas cools, and then suffer fewer late mergers 
in the $\Lambda$-dominated simulations.
It is also likely that some of the baryons in the inner parts
of the protogalaxies, where the specific angular momentum is 
lowest, either formed the luminous spheroids or were expelled 
entirely in galactic winds (Eke et al.\ 2000).

The specific angular momenta of real galactic disks can be
explained by a combination of extreme feedback and one or more 
of the other effects mentioned above.
Does this mean the angular-momentum problem has been solved? 
Yes and no.
Yes, because we now know that some mechanisms are capable of
reconciling the simulations with observations.
No, because we do not yet know whether these mechanisms operate 
in real galaxies.
The issue of feedback is especially vexing because it probably
depends on features of the interstellar medium on the scales of 
individual star-forming clouds, of order parsecs. (For a promising 
model of feedback, see Efstathiou 2000.)
The energy released by young stars---in ionizing radiation, stellar 
winds, and supernova ejecta---is more than sufficient to alter 
radically the collapse of the baryons in protogalaxies.
The key issues are where, when, and how this energy is deposited,
in particular, whether it is absorbed within the clouds themselves
or is distributed more widely within the protogalaxies. 
This, in turn, depends on the sizes and masses of the clouds,
the locations of the stars within them, whether the clouds are 
porous, and so forth.

\section{Perspective}

The goal of the research reviewed here has been to find a 
physically consistent model that accounts for the observed 
properties of galactic disks.
Two decades ago, it seemed as if we might be close to achieving
this goal.
At that time, the observed properties of galactic disks were
explained remarkably well by an idealized model in which the 
baryons and dark matter in protogalaxies acquired the same 
specific angular momenta by tidal torques and then conserved 
them during and after the collapse of the baryons.
A decade later we recognized that, if the gas in protogalaxies
were to cool rapidly, most of the angular momentum of the baryons 
would be transferred to the dark matter.
We now know that this angular-momentum problem can be solved 
by stellar feedback and other effects in principle, but we do
not yet know how effective these processes are in practice.
In another decade, with some luck, NGST will be in operation,
and we should be able to observe the formation of galactic disks
directly. 
It will be interesting to see how much theoretical and numerical
progress has been made by then.
The simulations will undoubtly improve, but it will be a formidable
challenge to include all the relevant physics on all the relevant
scales.

\bigbreak

\noindent {\bf Appreciation.} 
This article is dedicated to Ken Freeman at the celebration of 
his sixtieth birthday on Dunk Island. I am grateful to Ken for 
his friendship and collaboration over many years.

\section*{Discussion}

\noindent {\it Quinn:\,}
With respect to the angular-momentum catastrophe, I think 
we all knew the simulations had problems since the cooling
mechanisms (prescriptions) were not well defined physically
and the heating was totally absent.
Secondly, the loss of $J/M$ by a factor $\sim 10$ is a great
success in making ellipticals and spheroids, as pointed out by 
Zurek, Quinn, \& Salmon (1988).

\noindent {\it Fall:\,}
I tend to agree with both of your points.
Thanks for reminding me of your paper with Zurek and Salmon.
My impression is that the simulations without stellar feedback 
produce mainly spheroid-like objects, whereas those with extreme
feedback produce mainly disk-like objects.
Eventually, we should aim to produce both types of objects in 
the observed proportions in the same simulations.
It would be nice to find a simple mechanism to do this.

\noindent {\it Silk:\,}
The feedback models in which the gas is kept hot until $z \sim 1$
are going to have a problem in accounting for the observations of
relatively massive disks at $z \sim 1$ and also for the apparent
lack of evolution in the Tully-Fisher relation back to this
redshift.

\noindent {\it Fall:\,}
The kinds of observations you mention are potentially valuable 
constraints on, or input to, the models.
However, my impression is that such comparisons are still very
tentative.
The samples of galaxies at high redshifts have not been selected
in the same way as the nearby samples, and the photometric evolution
of the disks remains quite uncertain.

\noindent {\it Illingworth:\,}
Physically, why is it that $\Lambda$-dominated cosmologies help
with the angular-momentum problem?

\noindent {\it Fall:\,}
In $\Lambda$-dominated models, protogalaxies acquire their
angular momenta earlier, before the gas cools, and are then
disrupted by fewer late mergers than in matter-dominated 
models (see Eke et al.\ 2000).

\noindent {\it Bosma:\,}
Are the angular-momentum and cuspy-halo problems related?

\noindent {\it Fall:\,}
One might think so. 
Both problems seem to be aggravated by substructure within 
galactic halos.
And for this reason, they should both be alleviated in models 
with warm dark matter.
Recent simulations indicate, however, that WDM has more impact 
on the angular-momentum problem than it does on the cuspy-halo 
problem.

\end{document}